\documentclass [aps, prl, amssymb, twocolumn, showpacs,superscriptaddress]{revtex4-1}

\usepackage{graphicx,graphics,color,epsfig}
\usepackage{amsmath}
\usepackage{amssymb}
\usepackage{amsfonts}
\usepackage{amsfonts}
\usepackage{epstopdf}
\usepackage{bm}
\usepackage{times,xspace}
\usepackage[colorlinks,citecolor=blue,linkcolor=red]{hyperref}
\usepackage{color}

\begin{document} 
 
\title{Heat Diode Effect and Negative Differential Thermal Conductance across Nanoscale Metal-Dielectric Interfaces}

\author {Jie Ren}\email{renjie@lanl.gov}
\affiliation{Theoretical Division, Los Alamos National Laboratory, Los Alamos, New Mexico 87545, USA} 
\author{Jian-Xin Zhu}
\affiliation{Theoretical Division, Los Alamos National Laboratory, Los Alamos, New Mexico 87545, USA} 
\affiliation{Center for Integrated Nanotechnologies, Los Alamos National Laboratory, Los Alamos, New Mexico 87545, USA}


\date{\today}

\begin{abstract}
{Controlling heat flow by phononic nanodevices has received significant attention recently, because of its fundamental and practical implications. Elementary phononic devices such as thermal rectifiers, transistors and logic-gates are essentially based on two intriguing properties: heat diode effect and negative differential thermal conductance. However, little is known about these heat transfer properties across metal-dielectric interfaces, especially at nanoscale. Here we analytically resolve the microscopic mechanism of the nonequilibrium nanoscale energy transfer across metal-dielectric interfaces, where the inelastic electron-phonon scattering directly assists the energy exchange. 
We demonstrate the emergence of heat diode effect and negative differential thermal conductance in nanoscale interfaces and explain why these novel thermal properties are usually absent in bulk metal-dielectric interfaces. 
These results will generate exciting prospects for the nanoscale interfacial energy transfer, which should  have important implications in designing hybrid circuits for efficient thermal control and open up potential applications in thermal energy harvesting with low-dimensional nanodevices.} 
\end{abstract}


\maketitle

In metal, electrons dominate heat conduction, whereas in dielectric phonons are main energy carriers \cite{Mahanbook, Ziman}.
Hence, for metal-dielectric interfacial heat transport, energy transfer includes three possible pathways: (1) energy exchange between phonons in metal and phonons in dielectric, which is widely studied with the acoustic and diffuse mismatch models \cite{CahillJAP03, AMM2, DMM,PReddy:2009}; 
(2) nonequilibrium electron-phonon (e-ph) energy exchange within the metal, with subsequently phonon-phonon coupling across the interface \cite{Roukes85PRL, MajumdarAPL2004, JuJHT06};
(3) direct energy transfer from electrons in metal to phonons in dielectric through inelastic scattering induced by e-ph coupling across the interface \cite{eph1, eph2, eph3}.

When considering the pathway (2), one usually applies the phenomenological  two-temperature model \cite{TFM} that has been widely used to investigate the ultrafast pulsed laser heating of metals \cite{TFM_exp, TienCL}, and assume the pathway (3) is negligible \cite{MajumdarAPL2004, JuJHT06}. However, when the imperfect interface or the large lattice-mismatch of metal and dielectric is present, both pathway (1) and (2) will be seriously suppressed and the pathway (3) becomes significant. Recent experiments have demonstrated that when studying the metal-dielectric
interfacial heat transfer,  the e-ph coupling across the interface becomes crucial \cite{Hopkins1, Hopkins2}. In particular, the experiment in Ref. \cite{Guo_exp} reveals the strong direct coupling between electrons in metal and phonons in dielectric, which then dominates the overall interface heat transfer.  A particular challenge that arises is then to resolve the microscopic mechanism controlling the nanoscale interfacial energy transfer caused by the e-ph interaction across the metal-dielectric interface.

On the other hand, phononics recently emerges as a new discipline \cite{phononics}, where various functional thermal devices are designed to render the smart control of energy in micro-nano-scale. Among many intriguing properties, the heat diode effect \cite{diode1, diode2, diode22, diode3, diode4, diode5, diode6} and negative differential thermal conductance (NDTC) \cite{diode4, NDTC1, Segal06PRB, NDTC2, NDTC3, NDTC4, NDTC5} are the two most important ones for dielectric phononic devices acting as solid-state thermal rectifiers, switches, and  transistors etc~\cite{phononics}.
These dielectric phononic devices, when hybridized with metallic electronic circuits, then carry the potential of various beneficial applications, including the manipulation of heat dissipation and cooling in micro-nano-devices \cite{cooling, pop} and the renewable energy engineering, such as thermoelectric energy harvesting \cite{TE}. Therefore, understanding the interface heat transfer across nanoscale metal-dielectric hybrid structures is a long-standing challenge not only of fundamental interest, but also for practical applications in device design \cite{CahillJAP03, pop}. There are few previous efforts towards understanding the interfacial heat transfer in bulk metal-dielectric systems through the direct e-ph interaction  \cite{eph1, eph2, eph3}. In particular, the heat diode effect and NDTC have escaped from the attention,  and many important questions are left open: Can the bulk metal-dielectric with interfacial e-ph coupling  possess the heat diode and/or the NDTC properties? What if we scale the interface down to the micro-nano level? If so, what is their microscopic mechanism in the nanoscale interfacial energy transfer caused by the direct e-ph coupling? 

%
\begin{figure}
\rotatebox[origin=c]{0}{\includegraphics[width=0.7\columnwidth]{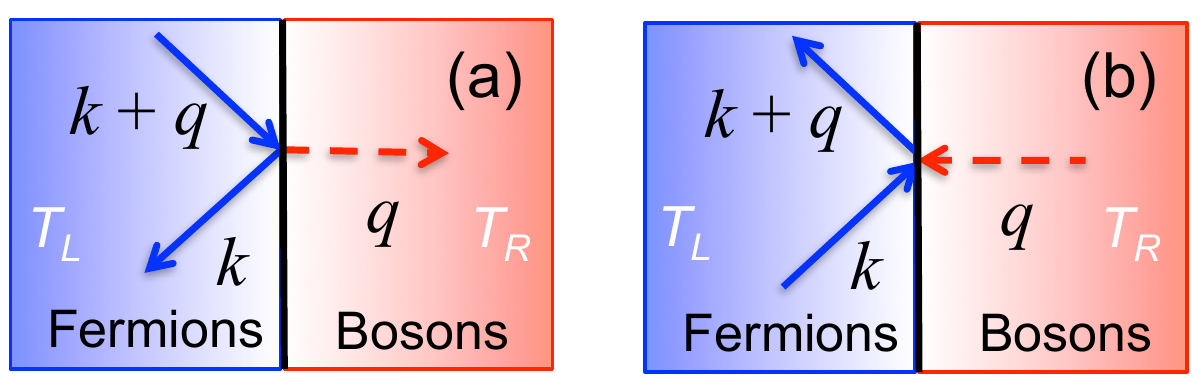}}
\vspace{-3mm} 
\caption{{\bf Sketch of inelastic scattering due to the e-ph interaction across the metal-dielectric interface.} (a) Down scattering of electrons with losing energy accompanied by phonon emission. (b) Up scattering of electrons with gaining energy accompanied by phonon absorption.} 
\label{fig1}
\end{figure}

In this work, we answer these above mentioned questions by analytically resolving the microscopic mechanism of the nonequilibrium nanoscale energy transfer across metal-dielectric interfaces, where the energy exchange is directly assisted by the interfacial e-ph interaction, i.e. the pathway (3) [see Fig.~\ref{fig1}]. We demonstrate that the heat diode effect and NDTC are usually absent from the macroscopic bulk metal-dielectric interfaces, and uncover that the non-smooth, strong energy dependent electronic density of state (DOS) is the crucial ingredient to retain these nontrivial properties. As such, we exemplify the identification of the heat diode effect and NDTC in several typical cases.
Our results describe the essential inelastic e-ph scattering around metal-dielectric interfaces in nanoscale and readily render clear physical insights and interpretations. These results could have important implication in providing the design principle for efficient hybrid thermal devices and open up potential applications in nanoscale thermal energy harvesting. 

\vskip0.25cm\noindent
{\bf Absence of heat diode and NDTC effects in macroscopic interfaces.}
Generally, the e-ph interaction around the metal-dielectric interface is described by the Hamiltonian \cite{Mahanbook, Ziman}:
 $H_{e-ph}=\sum_{kq} M_{kq} c^{\dag}_{k+q}c_ka_q+M^*_{kq} c^{\dag}_{k}c_{k+q}a^{\dag}_q$, 
where $c_k$ ($c^{\dag}_k$) annihilates (creates) an electron of momentum $k$, $a_q$ ($a^{\dag}_{q}$) annihilates (creates) a phonon with momentum $q$, and $M_{kq}$ is the e-ph coupling matrix. This Hamiltonian depicts two essential processes of inelastic e-ph scattering near the metal-dielectric interface that are responsible for the energy transport: i) an electron in the metal is down-scattered by emitting a phonon into the dielectric and the energy is transferred from the metal to the dielectric [see Fig.~\ref{fig1}(a)]; or ii) an electron in the metal is up-scattered by absorbing a phonon from the dielectric and the energy is transferred  from the dielectric back to the metal [see Fig.~\ref{fig1}(b)]. 

Starting form the Hamiltonian of e-ph coupling, by using the Fermi golden rule \cite{Mahanbook, eph1} or the standard Bloch-Boltzmann-Peierls formulas \cite{Ziman, Allen}, one can get the energy flux across the macroscopic interface as: 
\begin{widetext}
\begin{equation}
J=2\pi\int^{\infty}_0\!\!\!d\omega F_R(\omega) \hbar\omega \int^{\infty}_{-\infty} \!\!\! d\varepsilon \rho_L(\varepsilon) \left[f_L(\varepsilon+\hbar\omega)(1-f_L(\varepsilon))(1+N_R(\hbar\omega))-f_L(\varepsilon)(1-f_L(\varepsilon+\hbar\omega))N_R(\hbar\omega)\right],
\label{eq:bulkJ}
\end{equation}
\end{widetext}
which is from the electronic degrees of freedom (DOF)  with temperature $T_L$ on the left hand  side  to the  phononic DOF  with $T_R$ on the right hand side.
Here $f_{L}(\varepsilon)=[e^{{(\varepsilon-\mu_F)}/{(k_BT_L)}}+1]^{-1}$ is the Fermi-Dirac distribution for the electronic DOF, $N_{R}(\hbar\omega)=[e^{{\hbar\omega}/{(k_BT_R)}}-1]^{-1}$ is the Bose-Einstein  distribution for the phononic DOF, $\rho_L(\varepsilon)$ denotes the bulk electronic DOS and $F_R(\omega)$ contains the bulk phononic DOS and the e-ph coupling matrix elements~\cite{epf}. 
This expression is familiar from earlier discussions of energy transfer between electrons and phonons either within a single system \cite{Mahanbook, Allen} or between separate systems \cite{eph1, eph2, eph3}. To further proceed, one usually considers a good metal, i.e., the integral over $d\varepsilon$ converges within a thermal energy $k_BT_L$ around the Fermi level $\mu_F$ so that the bulk electronic DOS can be taken as a constant $\rho_L(\varepsilon)\approx \rho_L(\mu_F)$. Then, by applying the equality $\int d\varepsilon f_L(\varepsilon+\hbar\omega)(1-f_L(\varepsilon))=\hbar\omega/[e^{(\hbar\omega)/(k_BT_L)}-1]^{-1}=\hbar\omega N_L(\hbar\omega)$, one finally arrives  \cite{Mahanbook, eph2, eph3, Allen} at a Landauer-like formula, 
\begin{equation}
J\!=\!2\pi\rho_L(\mu_F)\!\!\int^{\infty}_0\!\!\!\!d\omega F_R(\omega) (\hbar\omega)^2\! \left[N_{\!L}(\hbar\omega)\!-\!N_{\!R}(\hbar\omega)\right].
\label{eq:Landauer}
\end{equation}
Since in the Landauer-type formula the temperature dependence only manifests  in the distribution difference $N_L(\hbar\omega)-N_R(\hbar\omega)$, it is straightforward to verify that this macroscopic bulk interfacial system can never have the heat diode and NDTC effects. This may also explain why the heat diode effect is absent in a bulk copper-copper oxide system \cite{Copper}, because cooper is a good metal.

\vskip0.25cm\noindent
{\bf Emergence of heat diode and NDTC effects by engineering the electronic DOS.}
From the above analysis, we see that the constant bulk DOS is the crucial assumption leading to the absence of heat diode and NDTC effects. 
However, when $\rho_L(\varepsilon)$ is not a smooth function but varies strongly on energy, which is exactly the case in low-dimensional nanoscale systems or can be achieved by material design and engineering,  we have to keep this electronic DOS inside the energy integral. In this way, the intriguing properties of heat diode effect and NDTC are able to emerge. 

\begin{figure}[!htb]
\vspace{-4mm}   
\rotatebox[origin=c]{0}{\includegraphics[width=1.0\columnwidth]{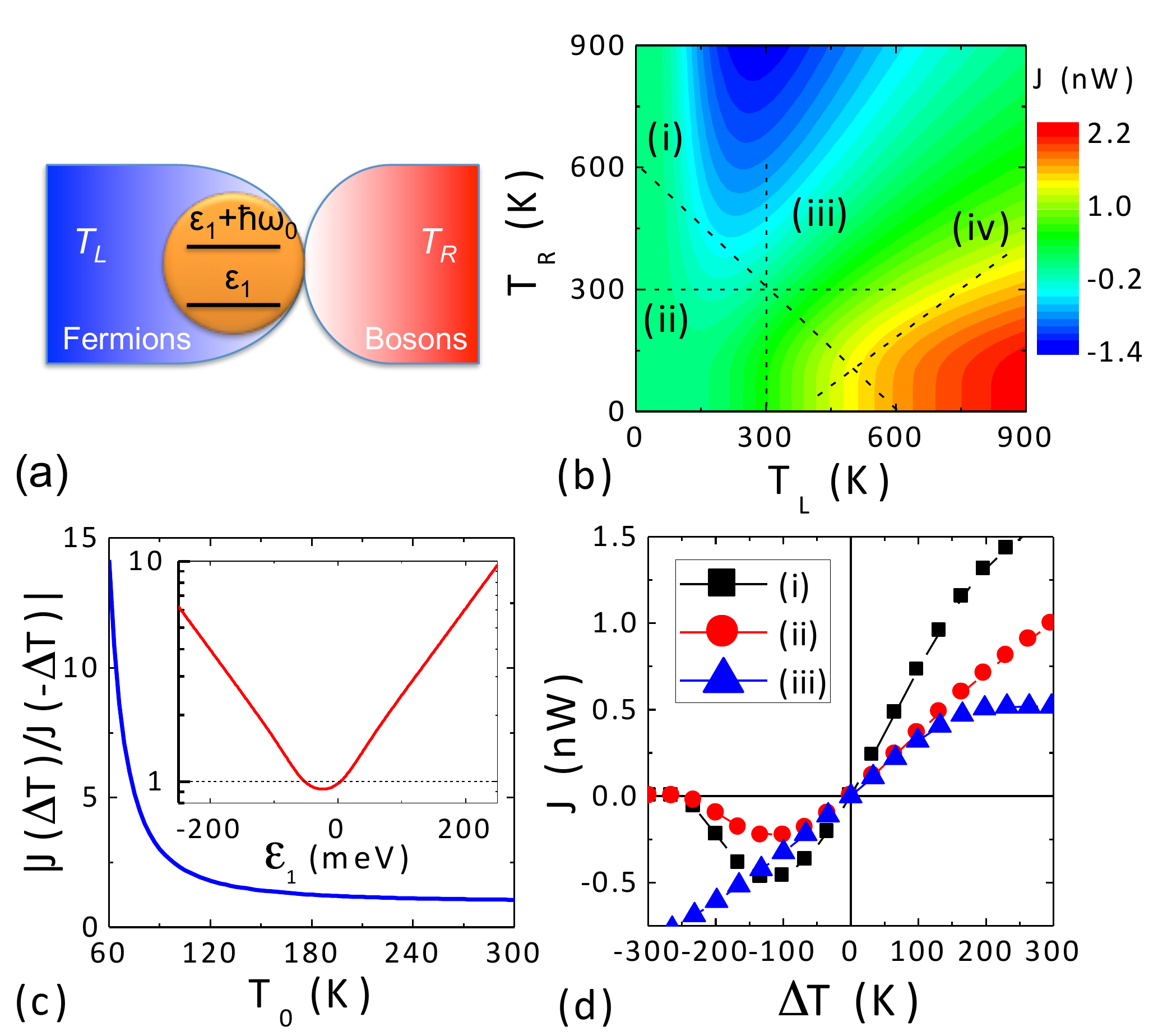}}
\vspace{-7mm} 
\caption{{\bf Heat diode effect and NDTC in the two-level interface system.} (a) Sketch of the two-level system setup describing the
nanoscale metal-dielectric interface. (b) Heat flux $J$ as a function of the fermionic temperature $T_L$ and the bosonic temperature $T_R$. Parameters are $\Gamma_L=\Gamma_R=2.5$ meV$/\hbar$, $\varepsilon_1=25$ meV, $\omega_0=50$ meV. (c) The rectification ratio as a function of $T_0$ for $T_L=T_0\pm\Delta T/2$, $T_R=T_0\mp\Delta T/2$ with $\Delta T=30$ K. Inset depicts the rectification ratio as a function of the lower level energy at $T_0=200$ K and $\Delta T=30$ K. Other parameters are the same as above.  (d) Heat flux vs $\Delta T$ along three corresponding lines in (b): (i) $T_L=T_0+\Delta T$, $T_R=T_0-\Delta T$; (ii) $T_L=T_0+\Delta T, T_R=T_0$; (iii) $T_L=T_0, T_R=T_0-\Delta T$, with $T_0=300$ K.
} \label{fig2}
\end{figure}

To demonstrate this point, let us assume the interface system is confined in two-dimension (2D) so that the left electronic part is modeled as 2D free electron gas which has a constant DOS beyond a minimum energy. By tuning $\mu_F$ to the bottom of the quadratic dispersion, we have $\rho_L(\varepsilon)=\rho_L(\mu_F)\Theta(\varepsilon-\mu_F)$. Thus, from Eq. (\ref{eq:bulkJ}) one arrives at
\begin{eqnarray}
J=2\pi\rho_L(\mu_F)\int^{\infty}_0d\omega  \mathcal T(\omega) \left[N_L(\hbar\omega)-N_R(\hbar\omega)\right],  \quad \label{eq:noLandauer}
\end{eqnarray}
where $ \quad \mathcal T(\omega)=F_R(\omega) \hbar\omega k_BT_L\ln{\frac{2e^{\hbar\omega/(k_BT_L)}}{e^{\hbar\omega/(k_BT_L)}+1}}$. 
As such, the temperature dependence of $\mathcal T(\omega)$ makes the heat diode and NDTC effects possible in this 2D metal-dielectric interface system. 
In fact, the heat flux in Eq.~(\ref{eq:noLandauer}) has qualitatively the same pattern of temperature dependent $J$ as in Fig. \ref{fig2}, which we will discuss in detail below. The 1D case will have the similar properties since its electronic DOS has highly fluctuating energy dependence.

One can also  consider a Lorentzian-type DOS $\rho_L(\varepsilon)=\frac{1}{\pi}\frac{W}{(\varepsilon-\varepsilon_0)^2+W^2}$, which is quite common in various physical systems. One of the famous physical realizations is a local state acting as a resonant level immersing in the continuum bulk conducting states, the so-called Fano model \cite{Fano}. When the half-width $W$ is small, the resonant peak of DOS becomes sharp at $\varepsilon_0$, so that from Eq. (\ref{eq:bulkJ}) one obtains
\begin{eqnarray}
J=2\pi\int^{\infty}_0d\omega F_R(\omega) \hbar\omega  I(\omega),
\label{eq:Lorenzian}
\end{eqnarray}
where $I(\omega)=f_L(\varepsilon_0+\hbar\omega)[1-f_L(\varepsilon_0)][1+N_R(\hbar\omega)]-f_L(\varepsilon_0)[1-f_L(\varepsilon_0+\hbar\omega)]N_R(\hbar\omega)$. This expression is familiar from Eq. (\ref{eq:heatflux}) which we will discuss in detail soon for a two-level interface. The difference is merely that instead of a single-mode phonon transfer there, the energy flux here has an integral over all phononic spectrum. This case also has the heat diode and NDTC effects, similar to the behaviors of Eq. (\ref{eq:heatflux}) that will be displayed in Fig. \ref{fig2}. In fact, this Lorentzian DOS example as well as the following two-level setup can be regarded as 0D interfaces whose electronic states are highly localized and thus possess sharp peaked DOS.

\vskip0.25cm\noindent
{\bf Heat diode effect and NDTC across a two-level interface.} When the feature size of electronic and phononic devices continues to scale down, as in the situation of quantum point contacts, the possible channels for  the inelastic e-ph scattering will be severely limited. The minimal interface containing the two essential processes of the scattering can be described by a  typical two-level system: $\varepsilon_1d^{\dag}_1d_1+\varepsilon_2d^{\dag}_2d_2$, where $\varepsilon_{1,2}$ denote two local energy levels with a gap $\hbar\omega_0=\varepsilon_2-\varepsilon_1$ [see Fig. \ref{fig2}(a)]. Two-level
system is a typical physical system that has many realizations in diverse fields, including quantum dots, single-electron-transistors, atomic-optical cavities, molecular junctions \cite{Galperin} with HOMO and LOMO levels and so on. The e-ph interaction at the interface between the left electronic reservoir and the right phononic reservoir is described by $\sum_qM_{q}d^{\dag}_2d_1a_q+M^*_{q}d^{\dag}_1d_2a^{\dag}_q$, which assists the relaxation and excitation of the two-level system and is responsible for the energy exchange.
In this case, it is readily to obtain the energy flux across the two-level interface \cite{supple}:
\begin{equation}
J\!=\!\hbar\omega_0\Gamma_{\!L}\Gamma_{\!R}\frac{f_{\!L2}\left(1\!-\!f_{\!L1}\right)\left(1\!+\!N_{\!R}\right)\!-\!f_{\!L1}\left(1\!-\!f_{\!L2}\right)N_{\!R}}{C},
\label{eq:heatflux}
\end{equation}
where $f_{Ln}=[e^{{(\varepsilon_n-\mu_F)}/{(k_BT_L)}}+1]^{-1}$ are the Fermi-Dirac distributions on the two-level interface, $N_R=[e^{{\hbar\omega_0}/{(k_BT_R)}}-1]^{-1}$ is the Bose-Einstein distribution for the phonons, $\Gamma_{L,R}$ are the corresponding tunneling rates to the electron and phonon parts, and $C=\Gamma_R\left[\left(1+f_{L2}\right)\left(1+N_R\right)+\left(1+f_{L1}\right)N_R\right]+\Gamma_L\left(1-f_{L1}f_{L2}\right)$. 
As we can see, the energy flux $J$ is proportional to the difference between two rate products: $f_{L2}\left(1-f_{L1}\right)\left(1+N_R\right)-f_{L1}\left(1-f_{L2}\right)N_R$. 
The first product $f_{L2}\left(1-f_{L1}\right)\left(1+N_R\right)$ describes the relaxation rate of the two-level interface from the occupied higher level to the empty lower one with releasing a phonon with energy $\hbar\omega_0$ into the right phononic lead. The second product $f_{L1}\left(1-f_{L2}\right)N_R$ reversely describes the excitation rate of the two-level interface from the occupied lower level to the empty higher one with absorbing a phonon with energy $\hbar\omega_0$ from the right part. 

Set  $T_L=T_0+\Delta T/2$, $T_R=T_0-\Delta T/2$, the heat flux can be generally expanded as $J=\kappa \Delta T+D \Delta T^2 + O(\Delta T^3)$, where the nonzero coefficient $D$ as a result of the system asymmetry is responsible for the heat diode effect. When $\partial J/\partial \Delta T=\kappa+D\Delta T+O(\Delta T^2)<0$, we say we have NDTC. In fact, from Eq.~(\ref{eq:heatflux}), we can verify that when fixing $T_L$, $\partial J/\partial \Delta T=-\partial J/\partial T_R$ is always positive, i.e., we can never have NDTC when we merely vary the temperature of the bosonic bath. For merely varying the fermionic bath temperature $T_L$, we find that $\partial J/\partial \Delta T=\partial J/\partial T_L>0$ under the condition $\varepsilon_1+\omega_0 \geqslant \mu_F \geqslant \varepsilon_1$. Therefore, the NDTC is only possible when we at least vary the temperature of the left electronic lead and when the two levels at the interface are both either above or below the Fermi energy $\mu_F$ of the left electrode.

%

Without loss of generality, we set the Fermi energy $\mu_F=0$ as the energy reference point. 
We plot a typical heat flux pattern as a function of both $T_L$ and $T_R$ in Fig.~\ref{fig2}(b). It is seen that the heat flux pattern is asymmetric with respect to the separatrix $T_L=T_R$, which demonstrates clearly the heat diode effect. 
For brevity, we just set the symmetric $\Gamma_L=\Gamma_R$. This emphasizes the intrinsic asymmetries inherent in the e-ph coupling across the metal-dielectric interface, while in previous studies of different systems, the heat diode effect requires asymmetric $\Gamma_L\neq\Gamma_R$, e.g., Ref.~\cite{diode5}.
The rectification ratio $|J(\Delta T)/J(-\Delta T)|$ is defined to quantify the asymmetry of energy transfer under opposite temperature bias. As shown in Fig.~\ref{fig2}(c), the heat diode effect becomes significant when temperature is cooled down and the asymmetric heat transfer prefers the direction from electrons in metal to phonons in dielectric. In this way, the device acts like a good thermal conductor while in the reversed way, the device acts like a thermal insulator.  The inset shows that the diode effect can be enhanced by tuning the level energy apart from the Fermi level $\mu_F=0$. Moreover, it shows that the diode can be reversed near the Fermi level so that the asymmetric energy transfer prefers the direction from phonon in dielectric to electrons in metal, although this reverse effect is not distinct. Of course, enlarging the temperature difference can also enhance the diode effect due to its nonlinear response nature.

For the behavior of the other nonlinear response quantity, i.e., differential thermal conductance (DTC), Fig.~\ref{fig2}(d) shows that for merely varying $T_R$ [see the line (iii)], the system always exhibits a positive DTC. Only with varying $T_L$ [see the line (ii)] or both $T_L, T_R$ [see the line (i)], the NDTC emerges. This can be understood through Eq.~(\ref{eq:heatflux}): 
In the regime of $T_R>T_L$, when $T_L$ approaches to the lower temperature regime, both $f_{L1}, f_{L2}$ approach to $0$ if $\varepsilon_1+\omega_0>\varepsilon_1>0$ or $f_{L1}, f_{L2}\rightarrow 1$ if $0>\varepsilon_1+\omega_0>\varepsilon_1$. In this way, the relaxation and excitation processes due to the e-ph interaction are effectively suppressed so that the energy flux $|J|$ deceases to zero, although the thermal bias $T_R-T_L$ increases.  

In the regime of $T_L>T_R$, we can also have the NDTC effect if varying both temperatures. In fact, when $T_L$ is large enough and $T_R$ is low enough, we have $f_{L1}\approx f_{L2}\approx 1/2$ and $N_R\approx 0$. In this way, $J\approx{\frac{1}{4}\hbar\omega_0\Gamma_L\Gamma_R}/({\frac{3}{4}\Gamma_L+\frac{3}{2}\Gamma_R})$. Now increasing $T_R$ and the thermal bias $T_L-T_R$ [e.g., see the line (iv) in Fig.~\ref{fig2}(b)], $f_{L1}, f_{L2}$ will fix at the half value and $N_R$ becomes some finite positive number. This modulation does not change the numerator of $J$ but increases the denominator, as: $J\approx{\frac{1}{4}\hbar\omega_0\Gamma_L\Gamma_R}/({\frac{3}{4}\Gamma_L+\frac{3}{2}\Gamma_R+3\Gamma_RN_R})$, so that the energy flux deceases although the thermal bias increases. In fact, when $N_R\approx 0$, the relaxation and phonon emission is effective but the excitation and phonon absorption is suppressed. Thus, when $N_R$ increases, the excitation becomes effective and absorbs the phonon from the right phononic bath through the e-ph interaction which attenuates the energy transfer from metal to dielectric. Due to this physical reason, the NDTC appears in the regime of $T_L>T_R$ as well if both temperatures are varied.

The physical reason for the NDTC at metal-dielectric interfaces is different from the previous mechanism in dielectric solid systems where phonon-phonon interaction (anharmonicity, nonlinearality) is necessary to shift the phononic power spectrum \cite{phononics} or in quantum junctions where the strong limit of system-bath coupling is required \cite{Segal06PRB}. Here, the interface system is under the weak coupling and the NDTC is a consequence of the competition between relaxation/excitation processes due to the interfacial e-ph interaction.

We have shown that non-smooth, strong energy dependent electron DOS is crucial for the emergence of heat diode and NDTC effects, as in low-dimensional nanoscale systems.
In fact, the factor $F_R(\omega)$ containing the e-ph couplings is also generally electronic energy dependent \cite{Mahanbook}. By considering non-smooth, strong energy dependent $F_R(\varepsilon, \omega)$ in some designed interfacial systems, we also expect to have the diode and NDTC effects.

Finally, we note that our results can be extended to general fermionic-bosonic interfacial systems. For example, in the nanoscale metal-radiative field interfaces  with electron-photon couplings, or in the nanoscale metal-magnetic insulator interfaces with electron-magnon couplings~\cite{spinSeebeckdiode}, we expect to observe the similar effects. These results described here are relevant for understanding the interface energy transfer, which could have important implication in providing the design principle for efficient hybrid thermal devices and open up potential applications in nanoscale thermal energy harvesting.
Promising future directions include, introducing the local e-ph coupling within the left metal, integrating with phonon-phonon interactions between metal and dielectric, and combining with some {\it ab initio} electronic structure theory, like the density functional theory for more realistic calculations.

\begin{acknowledgments}
{This work was supported by the National Nuclear Security Administration of the U.S. DOE at LANL under Contract No. DE-AC52-06NA25396, and the LDRD Program at LANL (J.R.), and in part by the Center for Integrated Nanotechnologies, a U.S. DOE Office of Basic Energy Sciences user facility (J.-X.Z.).}
\end{acknowledgments}


\newpage

\setcounter{figure}{0}
\setcounter{equation}{0}

\renewcommand\thefigure{S\arabic{figure}}
\renewcommand\theequation{S\arabic{equation}}
\renewcommand\bibnumfmt[1]{[S#1]}
\renewcommand{\citenumfont}[1]{S#1}

\begin{center}
\Large{\textbf{Supplemental Materials}}
\end{center}

\begin{center}
\textit{for ``Heat Diode Effect and Negative Differential Thermal Conductance across Nanoscale Metal-Dielectric Interfaces''}
\end{center}

\vspace{0.5cm}

This supplement provides the detailed derivation of the two-level interface model, which readily leads to the expression Eq. (5) in the main text.

{\bf The Hamiltonian of the two-level interface system.}
This nanoscale metal-dielectric interface is described by the total Hamiltonian $H=H_S+H_L+H_R+V_L+V_R$, where the two-level system $H_S=\varepsilon_1d^{\dag}_1d_1+\varepsilon_2d^{\dag}_2d_2$ interacts with two leads $H_{\nu} (\nu=L, R)$ at different temperatures $T_{\nu}$ via the couplings $V_{\nu}$ [see Fig.~2(a) in the main text].
Two-level system is a typical physical system that has many realizations in diverse fields, including quantum dots, single-electron-transistors, atomic-optical cavities, molecular junctions with HOMO and LOMO levels and so on. 
The left electron reservoir is described by the free electron Hamiltonian: $H_L=\sum_k\varepsilon_kc^{\dag}_kc_k$,
with the corresponding left coupling term describing the exchange of electrons:
$\!\!V_L\!=\!\sum_k\lambda_{k1}\!\left(d_1^{\dag}c_k\!+\!c_k^{\dag}d_1\right)\!+\!\sum_{k'}\lambda_{k'2}\!\left(d_2^{\dag}c_{k'}\!+\!c_{k'}^{\dag}d_2\right)\!$. The level $1$ and $2$ are coupled to the different states with $k$ and $k'$. Hence there are no indirect coupling between the two levels caused by the reservoir. The right phononic reservoir is modeled by independent phononic modes: $H_R=\sum_q\hbar\omega_qa^{\dag}_qa_q$,
with the corresponding right coupling term describing the e-ph interaction that assists the relaxation and excitation of the two-level interface:
$V_R=\sum_qM_{q}d^{\dag}_2d_1a_q+M^*_{q}d^{\dag}_1d_2a^{\dag}_q$.

{\bf Deriving heat flux expression from quantum master equation.}
We consider the strong Coulomb-blockade regime so that the interface system has only three eigenstates: empty state $|0\rangle$, state occupying the lower level $|1\rangle$, state occupying the upper level $|2\rangle$, with relations $H_S|0\rangle=0$, $H_S|1\rangle=\varepsilon_1|1\rangle$, $H_S|2\rangle=\varepsilon_2|2\rangle$. We can release the Coulomb-blockade assumption  by considering an additional double-occupied state for finite Coulomb interaction and the results do not change qualitatively.  
Employing the quantum master equation in the interaction picture, the evolution of the total density matrix $\rho_{nm}=\langle n|\rho|m\rangle$, $(n,m\in0, 1, 2)$  is described by
$\!\!{d \rho_{nm}}/{dt}\!\!=\!\!-\frac{i}{\hbar}\left[V(t), \rho(0)\right]_{nm}\!\!-\!\frac{1}{\hbar^2}\!\!\int^t_0\!d\tau\left[V(t), \left[V(\tau), \rho(\tau)\right]\right]_{nm}$,
where $V(t)$ is given in the interaction picture $V(t)=e^{i(H_S+H_L+H_R)t}(V_L+V_R)e^{-i(H_S+H_L+H_R)t}$. Following the standard approach in open quantum systems \cite{BreuerBook}: 1) assuming the Born-Markov condition; 2) considering the system-bath coupling $V$ up to the second order; 3) applying the secular approximation so that the population dynamics is decoupled with the non-diagonal coherence dynamics,
we finally arrive at  the equation of motion for the three-state populations:
\begin{equation}
\!\!\!\left[\!\!
\begin{array}{c}
 \dot p_0 \\
 \dot p_1 \\
 \dot p_2   
\end{array}\!\!
\right]
\!\!=\!\!\left[
\begin{array}{ccc}
 \!\!\!-k_{0\!\rightarrow\!1}\!-\!k_{0\!\rightarrow\!2}\!\! &   k_{1\!\rightarrow\!0} & k_{2\!\rightarrow\!0}   \\
  k_{0\!\rightarrow\!1}& \!\!\!-k_{1\!\rightarrow\!0}\!-\!k_{1\!\rightarrow\!2} &  k_{2\!\rightarrow\!1} \\
  k_{0\!\rightarrow\!2}&  k_{1\!\rightarrow\!2} &\!\!\!\!-k_{2\!\rightarrow\!0}\!-\!k_{2\!\rightarrow\!1}\!\!
\end{array}
\right]\!\!\!
\left[\!\! 
\begin{array}{c}
 p_0 \\
 p_1 \\
 p_2   
\end{array}\!\!
\right]\!\!,\!\! \nonumber
\label{eq:EOM}
\end{equation}
as well as the heat flux from the left to the right:
$J=\hbar\omega_0\big(p_2 k_{2\rightarrow1}-p_1 k_{1\rightarrow2}\big)$,
where $p_n=\text{Tr}_{B}[\rho_{nn}]$ with $\text{Tr}_{B}[\cdot]$ tracing out the degrees of freedom of both baths and $k_{n\rightarrow m}$ denotes the transition rate from state $|n\rangle$ to $|m\rangle$. By imposing the steady state condition $\dot p_0(t)=\dot p_1(t)=\dot p_2(t)=0$ and the conservation law $p_0(t)+p_1(t)+p_2(t)=1$ we finally obtain the heat flux Eq. (5) in the main text. Note that if alternatively using the nonequilibrium Green's function method \cite{NEGFbook}, we will arrive at the same results as above in the weak system-bath coupling limit.

{\bf The microscopic expressions for the transition rates $k_{i\rightarrow j}$.}
The tunneling rates, describing the exchange of electrons between the left electron reservoir and the two-level system, are given by 
$k_{0\rightarrow n}=\frac{1}{\hbar^2}\sum_k\lambda^2_{kn}\int^{\infty}_{-\infty}d\tau e^{-i{\varepsilon_n}\tau/{\hbar}}\text{Tr}_{B}[c^{\dag}_k(\tau)c_k(0)] =\Gamma_L(\varepsilon_n) f_{Ln}$, 
$k_{n\rightarrow 0}=\frac{1}{\hbar^2}\sum_k\lambda^2_{kn}\int^{\infty}_{-\infty}d\tau e^{i\varepsilon_n\tau/{\hbar}}\text{Tr}_{B}[c_k(\tau)c^{\dag}_k(0)] =\Gamma_L(\varepsilon_n) (1- f_{Ln})$, 
where $n=1, 2$, $f_{Ln}=[e^{{(\varepsilon_n-\mu_F)}/{(k_BT_L)}}+1]^{-1}$ and 
$\Gamma_L(\varepsilon_n)=\frac{2\pi}{\hbar}\sum_k\lambda_{kn}^2\delta(\varepsilon_n-\varepsilon_k)=\frac{2\pi}{\hbar}\lambda_{\varepsilon_n}^2\rho_L(\varepsilon_n)$ with $\rho_L(\varepsilon)$ the temperature-independent density of state (DOS) of the left electronic bath. 

The relaxation/excitation rates of the two-level interface system due to the inelastic e-ph coupling,  accompanied by the phonon emission/absorption, are given by 
$k_{2\rightarrow1}=\frac{1}{\hbar^2}\sum_qM^2_{q}\int^{\infty}_{-\infty}d\tau e^{i\omega_0\tau}\text{Tr}_{B}[a_q(\tau)a^{\dag}_q(0)] = \Gamma_R(\omega_0) (1+N_R)$, 
 $k_{1\rightarrow 2}=\frac{1}{\hbar^2}\sum_qM^2_{q}\int^{\infty}_{-\infty}d\tau e^{-i\omega_0\tau}\text{Tr}_{B}[a^{\dag}_q(\tau)a_q(0)]=\Gamma_R(\omega_0) N_R$, 
where $N_R=[e^{{\hbar\omega_0}/{(k_BT_R)}}-1]^{-1}$, $\Gamma_R(\omega_0)=\frac{2\pi}{\hbar}\sum_qM^2_q\delta(\hbar\omega_0-\hbar\omega_q)=\frac{2\pi}{\hbar}M_{\hbar\omega_0}^2\rho_R(\hbar\omega_0)$ with $\rho_R(\hbar\omega)$ the temperature-independent DOS of the right phononic bath.


\begin{thebibliography}{300}
\bibitem{epf} Conventionally, the expression $\alpha^2F_R(\omega)$ is used, named as Eliashberg function. Here, we drop $\alpha^2$ for simplicity.
\end{thebibliography}
\end{document}